\documentclass[preprint,nofootinbib,aps,10pt,twocolumn,superscriptaddress]{revtex4} 
\usepackage[dvips]{graphicx}

\usepackage{array,amsmath,amsthm,feynmf}
\usepackage{bm}
\usepackage{slashed}
\usepackage{times}
\usepackage{epsfig}
\usepackage{graphicx}
\usepackage{color}
\usepackage{cancel}
\usepackage{amssymb}
\usepackage{textcomp}

\newcommand{\be}{\begin{equation}}
\newcommand{\ee}{\end{equation}}
\newcommand{\bea}{\begin{eqnarray}}
\newcommand{\eea}{\end{eqnarray}}

\begin{document}

\smash{\hspace{6 cm} NPAC-12-22}\vspace{-5cm}


\title{\Large Stepping Into Electroweak Symmetry Breaking: 
Phase Transitions and Higgs Phenomenology}

\vspace{4.0cm}
\author{Hiren H. Patel}
\email{hhpatel@wisc.edu}
\affiliation{
{University of Wisconsin-Madison, Department of Physics} \\
{1150 University Avenue, Madison, WI 53706, USA}}
\author{Michael J. Ramsey-Musolf}
\email{mjrm@physics.wisc.edu}
\affiliation{
{University of Wisconsin-Madison, Department of Physics} \\
{1150 University Avenue, Madison, WI 53706, USA}}
\affiliation{
{Kellogg Radiation Laboratory, California Institute of Technology}\\
{Pasadena, CA 91125 USA}}
\date{\today}
\begin{abstract}
We study the dynamics of electroweak symmetry-breaking in an extension of the Standard Model where the Higgs sector is augmented by the addition of a real ($Y=0$) isospin triplet.  We show that this scenario exhibits a novel, two-step electroweak phase transition, wherein the first step provides the strongly first order transition as required for electroweak baryogenesis followed by a second step to the Standard Model Higgs phase that also admits a cold dark matter candidate. We analyze the constraints on this scenario from recent results obtained at the Large Hadron Collider for the Higgs diphoton decay channel. We argue that this two-step scenario can be generalized to extensions of the Standard Model with additional higher-dimensional scalar multiplets that may yield realistic baryogenesis dynamics.
\end{abstract}
\maketitle
\section{Introduction}
An outstanding problem at the interface of  cosmology with high-energy and nuclear physics is to explain the origin of baryon asymmetry of the universe:
\be
\eta=\frac{n_B}{n_\gamma}=( 5.54\pm 0.06)\times 10 ^{-10} \enspace\enspace\text{Planck\cite{Ade:2013zuv}}
\ee
where $n_B$ and $n_\gamma$ are the baryon and photon densities, respectively, and where the value has been determined from studies of the cosmic microwave background. Assuming a matter-antimatter symmetric initial conditions, three ingredients \cite{Sakharov:1967dj} must have been present in the microphysics of the early universe to generate a non-vanishing $\eta$: (1) violation of baryon number; (2) violation of both C- and CP-symmetry; and (3) departure from equilibrium dynamics or CPT violation. 
An attractive mechanism for solving this problem is electroweak baryogenesis (EWB) (for a recent review and extensive references, see Ref.~\cite{Morrissey:2012db}), wherein baryon number generation is driven by the generation of CP asymmetry at the time of the electroweak phase transition (EWPT).  An important ingredient for its success is the existence of a strong first order electroweak phase transition in which electroweak symmetry-breaking (EWSB) proceeds via bubble nucleation.

It is well known that the scalar sector of the minimal Standard Model (SM) is unable to have generated a strong first order phase transition because the Higgs boson is too heavy\footnote{The EWPT in the SM appears to be a cross-over transition\cite{Morrissey:2012db}}.  Extensions of the scalar sector of the Standard Model have been motivated, in part, by alleviating this difficulty.  Approaches include (a) the introduction of new scalars that increase the  barrier between the broken and unbroken phases associated with the non-analytic $\mathcal{O}(T)$ in the finite temperature effective potential; (b) the introduction of tree-level cubic terms in the $T=0$ potential that yield a barrier between the two phases; and (c) reliance on logarithmic corrections to the potential that can drive the transition (for a recent discussion, see Ref.~\cite{Chung:2012vg}). 


Most analyses of the EWPT have thus far relied on only one of these approaches while making the reasonable -- but not necessary -- assumption that EWSB proceeds in a single step. It is possible, however, that the dynamics of EWSB are more complicated, involving more than one of the aforementioned mechanisms and proceeding through multiple steps, passing through intermediary phases before reaching the EW phase.  In this work, we study a minimal scalar sector extension that gives rise to a two-step transition and that relies on two of the general mechanisms listed above. Specifically, we add a real triplet $\vec\Sigma$ that transforms as $(1,3,0)$ under  $\text{SU(3)}_\text{C}\times\text{SU(2)}_\text{L}\times\text{U(1)}_\text{Y}$, corresponding to the smallest dimension scalar multiplet that carries non-trivial $\text{SU(2)}_\text{L}$ charge. Using this minimal scenario, we show how a multiple step transition may facilitate the first order EWPT needed for EWB,  yield a dark matter particle, and contain dynamics testable through Higgs boson decays into diphotons at the Large Hadron Collider.

\begin{figure}
\includegraphics[scale=1.0,width=6cm]{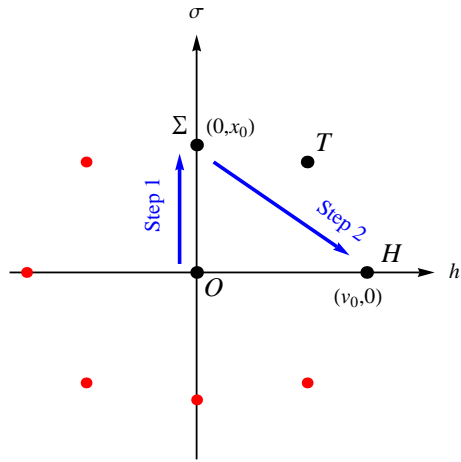}
\caption{Field phase space indicating critical (extremal) points in the tree-level potential at zero temperature, and the expected two-step pattern of symmetry breaking at finite temperature.  Red points are related to black points via $(Z_2)_H$ and $(Z_2)_\Sigma$ symmetries.}
\label{fig:vacuumSketch}
\end{figure}

In brief, at temperature $T=0$ the neutral component $\Sigma^0$ can provide a viable cold dark matter candidate when the scalar potential admits a $Z_2$ symmetry (for an analysis of the zero-temperature properties and collider phenomenology, see Ref.~\cite{FileviezPerez:2008bj}). At $T\sim 100$ GeV, the interplay of the scalar triplet and Higgs doublet fields (in the minimum of the free energy) may give rise to a two-step transition, schematically illustrated in Fig. \ref{fig:vacuumSketch}.  During the first step, at a temperature just below the first critical temperature $T_\sigma$, the system makes a transition from the symmetric phase at point $O$ to the isospin breaking phase at point $\Sigma$, where the neutral component of the triplet field obtains a non-vanishing vacuum expectation value (vev) $\langle \Sigma^0\rangle$, while the neutral Higgs vev remains zero.  Then, in the second step at a temperature below the second critical temperature $T_h$, the universe makes a transition from isospin breaking phase at point $\Sigma$ to the electroweak symmetry breaking phase at point $H$, where the Higgs vev $\langle H^0\rangle$ becomes non-vanishing but $\langle \Sigma^0\rangle$ relaxes to zero, ultimately leading to the $T=0$ Higgs phase with $\Sigma^0$ as the dark matter particle. 

While the remainder of the paper addresses the dynamics for this scenario in detail, we comment on several salient features here. 
\begin{itemize}
\item[(1)] Although the zero temperature $\Sigma^0$ vev in the Higgs phase need not vanish, the constraint from the electroweak $\rho$ parameter requires it to be small $\langle\Sigma^0\rangle < 4\text{ GeV}$. Allowing it to be tiny but non-vanishing does not substantially alter the EWPT dynamics but does preclude $\Sigma^0$ as a viable dark matter candidate. Consequently, we take $\langle \Sigma^0\rangle = 0 $ at the conclusion of the second step in order to yield a dark matter candidate. To that end, we impose a dark matter-preserving $(Z_2)_\Sigma$ symmetry on the potential and refer to the model as the ``$Z_2$$\Sigma$SM."
\item[(2)] The first step of the phase transition ($O\to\Sigma$) can be strongly first order, driven entirely by the finite-$T$ dynamics of the effective potential along the neutral triplet scalar direction. This transition is analogous to the one that might occur along the $O\to H$ direction in the Standard Model but is excluded from being first order due to non-existence of a sufficiently light Higgs boson. In contrast, the parameters in the $\vec\Sigma$ sector of the theory are sufficiently unconstrained by current phenomenology to allow for a strong first order EWPT along the  $O\to\Sigma$ direction. 
\item[(3)] Bubble nucleation during the first step ($O\to\Sigma$) creates the necessary environment for baryon number generation, assuming additional sources of CP-violation beyond those provided by the Standard Model. Since the $\vec\Sigma$-field carries non-trivial SU(2)$_L$ charge, the B+L-violating monopole interactions that destroy baryon number are suppressed inside the $\Sigma$-phase bubbles, capturing any net baryon number density produced ahead of the advancing bubble walls. In the present work, we concentrate on the phase transition dynamics for this step, leaving an analysis of possible sources of CP-violation to a future study.
\item[(4)] The dynamics of the second step ($\Sigma\to H$) are governed by the tree-level interaction between the $\vec\Sigma$ and $H$ fields. To ensure that any baryon number produced during the first step is not washed out by reactivation of the SM  sphalerons, this transition is also first order and sufficiently strong. Moreover, entropy production is not too copious without diluting the initial baryon asymmetry\footnote{We thank A. Kusenko for initial discussions of the latter point.}. As we discuss below, the degree to which these requirements are met can be constrained by measurements of the $H\to\gamma\gamma$ rate.
\item[(5)] While the introduction of the $\vec\Sigma$ field constitutes the minimal extension of the Standard Model scalar sector leading to this scenario, it is possible that scalars transforming under higher-dimensional isospin representations will yield similar dynamics, though not necessarily with a cold dark matter candidate as well. As we argue below, the main features are otherwise generic and are likely to persist in other models that also include additional degrees of freedom as needed for an appropriate ultraviolet completion. In this regard, a similar two-step scenario was considered\footnote{The authors thank M. B. Wise for alerting us to the existence of this work.} in \cite{Land:1992sm} involving a second Higgs doublet rather than a real triplet.  The authors envisioned baryon-asymmetry generation to occur during the much stronger second step of the phase transition.  However, as pointed out in \cite{Hammerschmitt:1994fn}, since weak isospin symmetry is already broken in the first step,  B+L violating processes are already too suppressed to convert existing CP asymmetry to baryon asymmetry. Hence, we concentrate on the case where the first step is strongly first order.
\end{itemize}


Our analysis of the foregoing scenario is organized as follows: in section \ref{sec:Model}, we formulate the model and subsequently discuss the zero temperature vacuum structure and tree-level vacuum stability constraints in section \ref{sec:vacStab}.  We then turn our attention to the finite-$T$ dynamics, focusing first on the $B+L$ violating interactions
in section \ref{sec:sphaleron} then following up with an analysis of the two-step EWPT in section \ref{sec:EWPT}.  Finally, we draw connections to collider phenomenology and discuss implications coming from recent LHC results in section \ref{sec:pheno}.

\section{Model and formulation}\label{sec:Model}
In the real triplet extension of the Standard Model, the scalar sector is composed of the standard Higgs doublet $H \sim (1,2,+1/2)$, and a real triplet\footnote{We follow the convention that the arrow above a letter denotes a vector in isospin space.} $\vec{\Sigma}\sim(1,3,0)$, with $Q_\text{EM}=T^3+Y$.  Our convention for component fields are
\begin{equation}\label{eq:fieldConv}
H=\begin{pmatrix}\phi^+\\\frac{1}{\sqrt{2}}(v_0+h+i\phi^0)\end{pmatrix},\qquad\vec\Sigma=\begin{pmatrix}\sigma_1\\\sigma_2\\\sigma_3+x_0\end{pmatrix}\,,
\end{equation}
where we fix the tree-level Higgs vacuum expectation value (VEV) at $v_0=246\text{ GeV}$ and where $x_0 = \langle \Sigma^0 \rangle$ is the triplet vev. We denote the physical quanta of charged and neutral scalar fields $\Sigma^\pm = (\sigma_1\mp i \sigma_2)/\sqrt{2}$ and $\Sigma^0=\sigma_3$, respectively.


Following the notation of our earlier work\cite{FileviezPerez:2008bj} we write the $Z_2$-symmetric scalar potential as 
\begin{align}\nonumber
V(H,\Sigma)&=-\mu^2 H^\dag H+\lambda(H^\dag H)^2-\frac{\mu_\Sigma^2}{2} (\vec\Sigma\cdot\vec\Sigma) \\
\label{eq:zeroTpotential}&\qquad+\frac{b_4}{4} (\vec\Sigma\cdot\vec\Sigma)^2+\frac{a_2}{2}H^\dag H (\vec\Sigma\cdot\vec\Sigma)\,.
\end{align}
This potential exhibits two independent global symmetries that are also accompanied by reflection symmetries\footnote{The Yukawa sector of the Standard Model Lagrangian breaks the reflection symmetry for the Higgs doublet.}:
\begin{align*}
\text{SO}(4)_H:&\enspace H\rightarrow O_{4\times 4}H&\hspace{4mm}(Z_2)_H:&\enspace H\rightarrow-H\\
\text{SO}(3)_{\Sigma}:&\enspace\vec\Sigma\rightarrow O_{3\times 3}\vec\Sigma&\hspace{4mm}(Z_2)_\Sigma:&\enspace \vec\Sigma\rightarrow-\vec\Sigma\,.
\end{align*}
We use these symmetries to restrict the discussion of the electroweak phase transition to electrically neutral components of the Higgs doublet and isospin triplet.  We should mention, however, that as detailed in \cite{FileviezPerez:2008bj}, the addition of a possible $a_1 H^\dag(\vec\Sigma\cdot\vec T)H$ breaks the $\text{SO}(3)_{\Sigma}$ symmetry and would trigger the acquisition of a vacuum expectation value for the neutral component of the isospin triplet, restricted to values of the order $\langle\Sigma^0\rangle\alt 3\text{ GeV}\ll\langle H\rangle=246 \text{ GeV}$ by the tightly constrained electroweak $\rho$-parameter at tree-level.  Although its presence may substantially affect collider phenomenology even when within bounds of experimental uncertainty for $\rho_\text{EW}$ through its mixing with the doublet, the impact on the electroweak phase transition is negligible.  Therefore, for clarity, we restrict our discussion of the phase transition to the region of the parameter space where the $(Z_2)_{\Sigma}$ symmetry is realized, thereby also allowing the neutral component to be a dark matter candidate.

This model contains only three real parameters in addition to those of the Standard Model: the triplet (negative) mass-parameter $\mu_\Sigma^2$, triplet self-coupling $b_4$, and the doublet-triplet \lq\lq Higgs-portal" coupling $a_2$.  The analysis is greatly simplified due to the addition of small number of undetermined parameters and the limited number of field degrees of freedom participating in the electroweak phase transition.

\section{Zero temperature vacuum structure}\label{sec:vacStab}
The qualitative behavior of the electroweak phase transition is largely influenced by the zero temperature vacuum structure of the Higgs sector.  In this section we perform a vacuum stability analysis at tree level.  Following our conventions in (\ref{eq:fieldConv}), we denote the neutral components of the isospin doublet and triplet as $h$ and $\sigma$ respectively.  Upon setting the remainder of the components to zero,
the potential reads 
\begin{align}\nonumber
V_\text{tree}(h,\sigma)&=-\frac{1}{2}\mu^2\,h^2-\frac{1}{2}\mu_\Sigma^2\sigma^2+\frac{1}{4}\lambda h^4\\
\label{eq:treePotential}
&\qquad+\frac{1}{4}b_4\sigma^4+\frac{1}{4}a_2\,h^2\sigma^2\,.
\end{align}
The critical points are found by solving the minimization conditions
\begin{equation}\label{eq:minCond}
\frac{\partial V_\text{tree}}{\partial h}\Big|_\text{crit}=\frac{\partial V_\text{tree}}{\partial \sigma}\Big|_\text{crit}=0\,.
\end{equation}

The potential, being a fourth-order polynomial in two variables, nominally has a total of nine extremal points.  Apart from the origin $(h,\sigma)=(0,0)$, the extrema come in positive/negative pairs related by combinations of the $(Z_2)_H$ and $(Z_2)_\Sigma$ reflection symmetries.  We use these symmetries to eliminate the redundant negative partners down to four distinct critical points as indicated by black points in Fig. \ref{fig:vacuumSketch}, labeled by $H$, $O$, $\Sigma$, and $T$.  We note that not all points may be realized as critical points of the potential as certain choices of parameters may yield complex-valued solutions (see potentials in Fig. \ref{fig:potentials}).

\begin{figure*}
\includegraphics[scale=1.0,width=7cm]{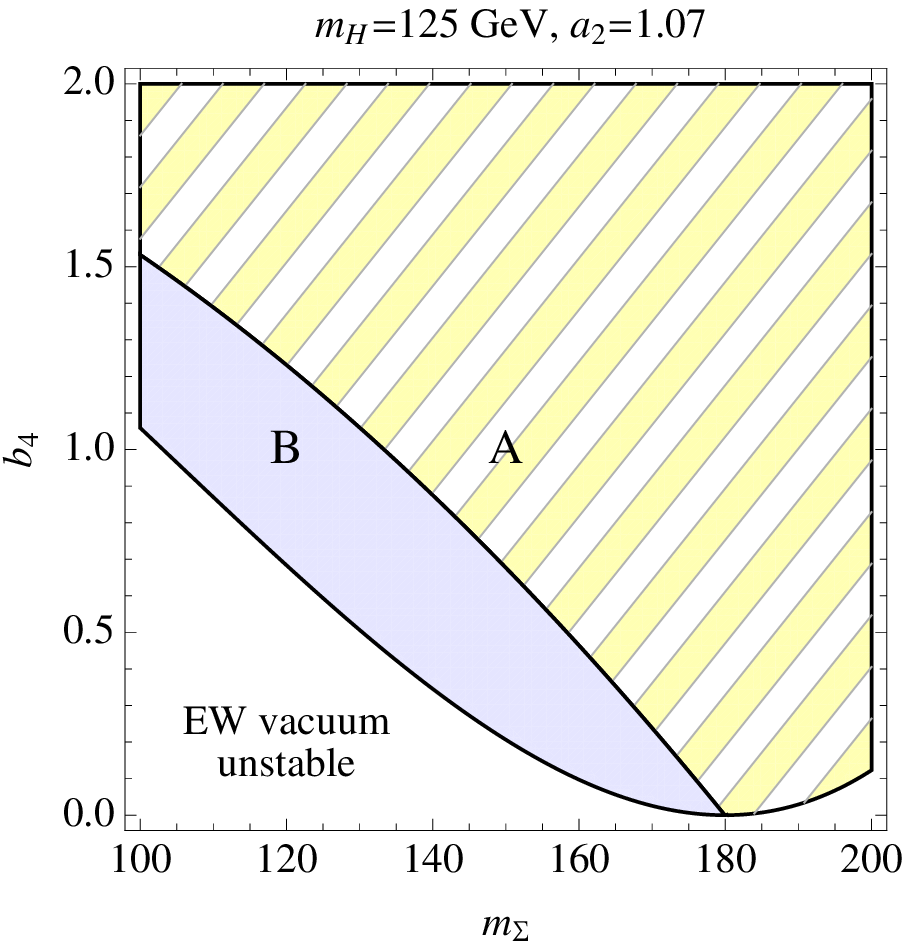}\qquad
\includegraphics[scale=1.0,width=7.25cm]{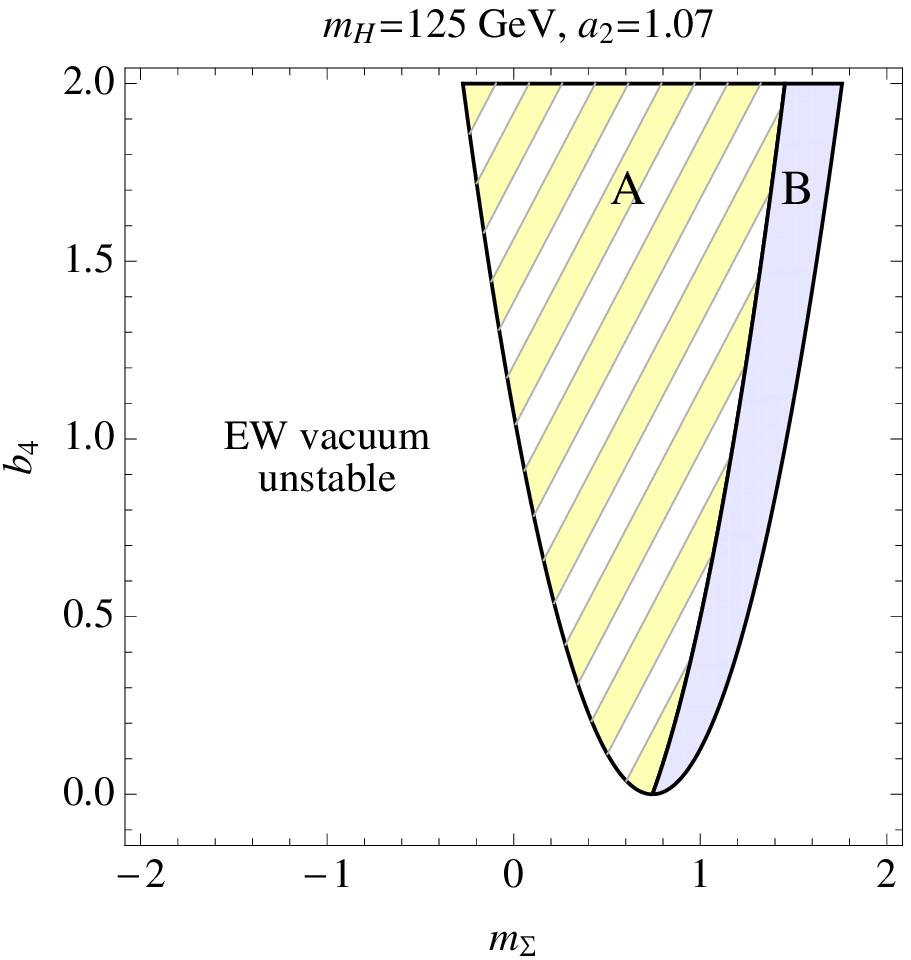}
\caption{Regions A (yellow striped) plus B (solid blue) indicate where tree-level electroweak vacuum stability condition  of (\ref{eq:explicitVacStab}) is satisfied. Left panel: the $m_\Sigma$-$b_4$ plane for fixed $(m_H=125\text{ GeV},\,a_2=1.07)$; right panel:  the $a_2$-$b_4$ plane for fixed $(m_H=150\text{ GeV},\,m_\Sigma=150\text{ GeV})$. Regions B indicate where (\ref{eq:metastableSigma}) is also satisfied and the tree-level potential exhibits a metastable minimum along neutral $\Sigma$ direction. Illustrative representations of the scalar potential for regions A and B are indicated in the left and right panels of Fig. \ref{fig:potentials}, respectively. }
\label{fig:vacuumStab}
\end{figure*}

\begin{figure*}
\includegraphics[scale=1.0,width=7cm]{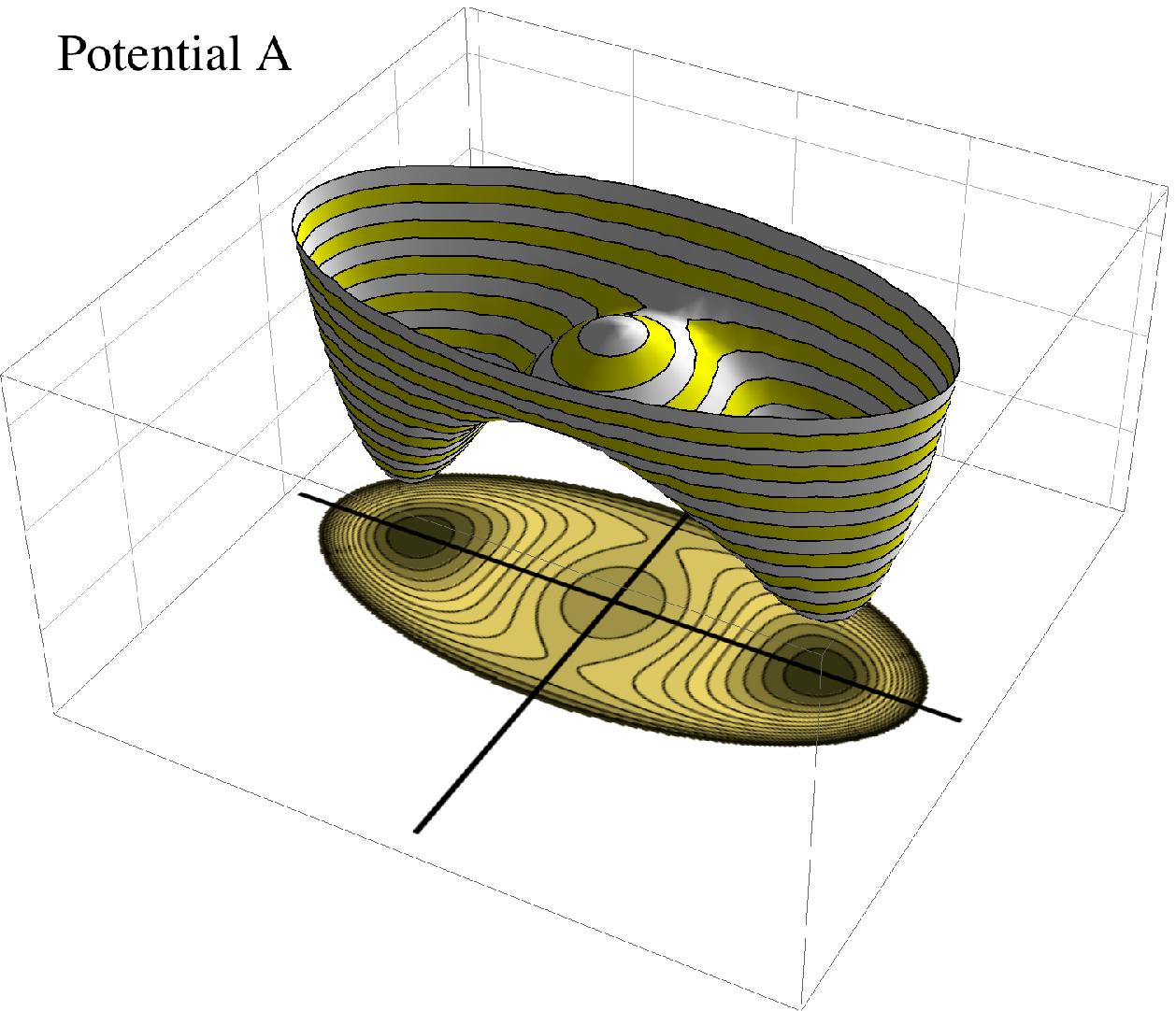}\qquad
\includegraphics[scale=1.0,width=7cm]{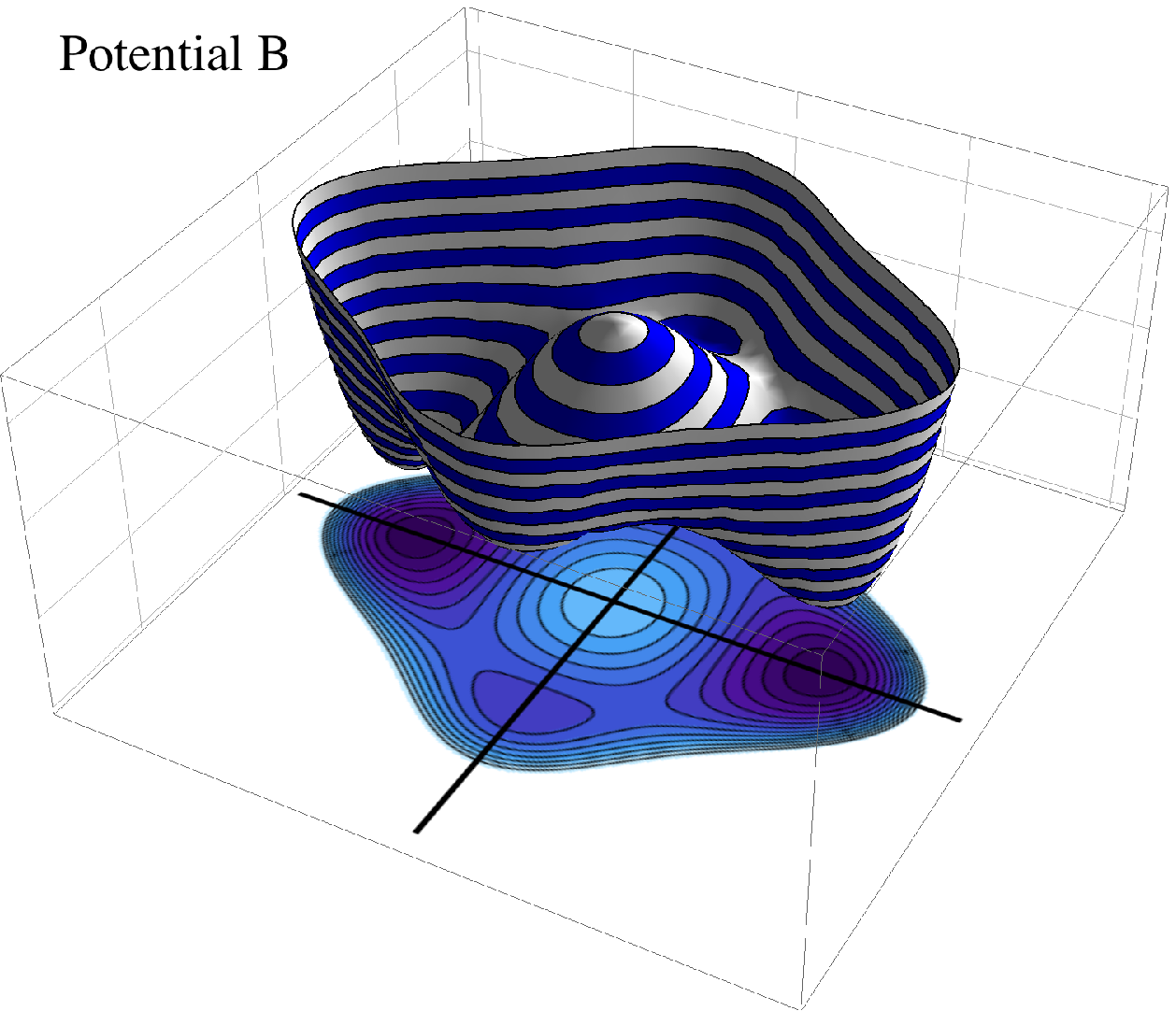}
\caption{Qualitative picture of the potential $V(h,\,\sigma)$ of (\ref{eq:treePotential}) in the two different regions of parameter space as indicated in Fig. \ref{fig:vacuumStab}.  Potential A (corresponding to regions A of Fig.~\ref{fig:vacuumStab})
displays no critical point along the $\sigma$ direction, whereas Potential B (corresponding to regions B of Fig.~\ref{fig:vacuumStab}) exhibits a metastable minimum along the $\sigma$-direction.}
\label{fig:potentials}
\end{figure*}

We label point $H$ as the Higgs phase point that  corresponds to the physical electroweak vacuum,  where the tree-level mass relations are determined to be
\begin{gather}
\begin{aligned}
m_H^2&=2\lambda v_0^2 \approx (125 \text{ GeV})^2\qquad\text{(LHC)}\\
\label{eq:msigma}
m_{\Sigma^\pm}^2=m_{\Sigma^0}^2&=-\mu_\Sigma^2+\frac{1}{2}a_2 v_0^2
\end{aligned}
\end{gather}
More generally, the requirements of vacuum stability may effectively be summarized by the condition that point $H$, the electroweak vacuum located at $(h,\,\sigma)=(246\text{ GeV},\,0)$, is the global minimum of the potential, with  $m_\Sigma^2>0$ and  $m_H^2>0$ masses.  This requirement is expressed by the inequality
\begin{equation}\label{eq:explicitVacStab}
\frac{1}{b_4}\Big(\frac{1}{2}a_2 v_0^2-m_\Sigma^2\Big)<\frac{1}{2}m_H^2 v_0^2\,.
\end{equation}

Further, in order to facilitate the discussion of two-step phase transitions, it will be useful to identify regions of parameter space where the potential exhibits a secondary local minimum at point $\Sigma$ with positive masses.  A straightforward calculation yields the condition for the existence of a secondary minimum,
\begin{equation}\label{eq:metastableSigma}
\frac{1}{2}m_H^2>\frac{1}{2}\frac{a_2}{b_4}\Big(\frac{1}{2}a_2v_0^2-m_\Sigma^2\Big)\,,
\end{equation}
which requires $\mu_\Sigma^2>0$ in (\ref{eq:msigma}).

In Fig. \ref{fig:vacuumStab} we display the region (shaded yellow and blue) in the $a_2$-$b_4$ plane for which the vacuum stability condition in (\ref{eq:explicitVacStab}) is satisfied, with the masses $m_\Sigma = 150$ GeV, and $m_H = 125$ GeV held fixed.  The blue shaded region indicates points where  the requirement (\ref{eq:metastableSigma}) is also satisfied and the potential has a secondary local minimum at point $\Sigma$. To assist the reader in visualizing the potential for various regions of parameter space, we provide illustrative plots in Fig. \ref{fig:potentials} of the potential for two cases:  (a) Eq. (\ref{eq:explicitVacStab}) only being satisfied corresponding to a representative point in the yellow region in Fig. \ref{fig:vacuumStab} and (b) both Eqs.~(\ref{eq:explicitVacStab}) and (\ref{eq:metastableSigma}) holding, corresponding to the blue region in Fig. \ref{fig:vacuumStab}.

\subsection{Quantum corrections}
Quantum corrections to the tree-level potential can, in general, have a significant impact on the vacuum structure of the theory. 
In the SM, for example, top quark fluctuations may cause the Higgs quartic self coupling $\lambda$ to run negative below the Planck scale if the input value at the electroweak scale -- set by the value of the Higgs mass -- is too small.  In the present instance, the potential remains stable along the $\sigma$-direction up to the Planck scale since the vanishing hypercharge of the $\vec\Sigma$ precludes any couplings to fermions and, thus, any substantial negative contributions to the $\beta$ function for the $\Sigma$ quartic self-coupling, $b_4$. Moreover, the $H^\dag H (\vec{\Sigma}\cdot\vec{\Sigma})$ interaction in  (\ref{eq:zeroTpotential}) generates a positive contribution to $\beta_\lambda$ proportional to $a_2^2$, partially compensating for the negative top quark contribution and improving stability of the Higgs and $\Sigma$ phases. On the other hand, $\lambda(\mu)$ may become non-perturbative below the Planck scale if $a_2$ is too large (for recent work on the implications for vacuum stability and perturbativity in similar scenarios for physics beyond the SM, see {\em e.g.} Refs.~\cite{Chao:2012mx,Chen:2012fa,Gonderinger:2012rd,Chun:2012jw,Cheung:2012nb,Lebedev:2012zw,EliasMiro:2012ay,Gonderinger:2009jp}).  Since our interest here focuses on the novel phase transition dynamics associated with this scalar sector extension, we defer to future work an analysis of the scale at which the $Z_2$$\Sigma$SM becomes non-perturbative.

\section{$B+L$ violation and Baryon Number Preservation Criterion}\label{sec:sphaleron}
Crucial to baryogenesis in this model is the two step process by which the phase transition proceeds.  In the first step the system makes a transition from the symmetric phase to the weak-isospin broken $\Sigma$ phase at a critical temperature we define as $T_\sigma$.  This step is to proceed via bubble nucleation since we expect this step to be the baryon-asymmetry generating step.  Subsequently, a second phase transition (at temperature $T_h$) to the familiar SM Higgs phase ($H$) occurs.  In this section we discuss sources of baryon number violating processes in each step that threaten to erase the asymmetry generated at the first step, and derive an approximate baryon number preservation criterion (BNPC) on the strength of each step of the EW phase transition.

\subsection{First Step: `t Hooft-Polyakov Monopoles}
Within the bubbles of the first step, the $\Sigma$ phase resembles that of the SU(2) Glashow-Salam \cite{Georgi:1972cj} model of electroweak interactions without weak neutral currents.  It was noticed some time ago \cite{'tHooft:1974qc,Polyakov:1974ek} that the model admits classical magnetic monopole solutions.  Studies by Rubakov \cite{Rubakov:1981tf} later revealed that fermion interactions with monopoles lead to fermion-number violation via the axial anomaly.  When embedded within the standard model, these monopoles are expected to violate baryon number via the B+L anomaly.

Within the context of our study, any monopoles inside the $\Sigma$ phase would, in principle, lead to a dilution of the baryon asymmetry generated at this step by an analogous reaction.  Consequently it is essential that the density of monopoles $n_\text{M}$ in these bubbles be sufficiently low to minimize the loss of baryon density.  There are two sources for the generation of magnetic monopoles in the $\Sigma$ phase.  Magnetic monopoles may be generated at bubble collisions (Kibble mechanism)\cite{Kibble:1976sj}, or by thermal monopole-antimonopole pair production.  In the former case, a lower bound for monopole density for a first order phase transition has been estimated in \cite{Preskill:1984gd}:
\begin{equation}\label{eq:monopoleKibb}
\frac{n_\text{M}}{T^3} \gtrsim p\left(\frac{T_\text{nuc}}{(0.6)n_\star^{-1/2}m_P}\right)^3\,,
\end{equation}
where $p\sim0.1$ is the probability that the scalar field orientation at a collision point is topologically non-trivial, $n_\star\sim 100$ is the effective number of relativistic degrees of freedom, $m_P\sim10^{19}\text{ GeV}$ is Planck mass, and $T_\text{nuc}\sim 10^{2} \text{ GeV}<T_\sigma$ is the temperature for bubble nucleation.  In the latter case, thermal monopole pair production will attempt to bring the monopole density to an equilibrium value of
\begin{equation}\label{eq:monopoleEquil}
\frac{n_\text{M}}{T^3}\Big|_\text{eq.} = \left(\frac{m_\text{M}(T)/T_\text{nuc}}{2\pi}\right)^{3/2}e^{-m_\text{M}(T)/T}\,,
\end{equation}
where $m_\text{M}(T)$ is the temperature-dependent monopole mass.  At the classical level, it is given by
\begin{equation}\label{eq:monopoleMass}
m_\text{M}({T})=\frac{4\pi \bar{x}(T)}{g}B_\text{M}(b_4/g^2)\,,
\end{equation}
where $B_\text{M}(b_4/g^2)$ is an $\mathcal{O}(1)$ function (see \cite{Forgacs:2005vx} for details), and $x$ is the value of the triplet field at the metastable point $\Sigma$ in Fig. \ref{fig:vacuumSketch}.  The temperature-dependent mass is derived by scaling\cite{Braibant:1993is} the vev based on the high-$T$ thermal potential in (\ref{eq:highT_V}) below,
$$x\rightarrow \bar{x}(T)\,.$$
Upon inserting (\ref{eq:monopoleMass}) into (\ref{eq:monopoleEquil}), quick numerics suggest that the equilibrium monopole density exceeds that due to bubble collisions (\ref{eq:monopoleKibb}) for all reasonable monopole masses.  However, the above analysis was carried out under the assumption that monopole-antimonopole thermal production rates are sufficiently fast to bring monopole density to thermal equilibrium.  Therefore, (\ref{eq:monopoleEquil}) represents an upper bound on monopole density.  

 The precise implications of the non-vanishing monopole density on baryon number washout require analysis of the rate equations for $(B+L)$-violating processes in the de-confined phase of QCD, as the temperature of the $\Sigma$ phase lies well above the confinement temperature. Such an analysis goes beyond the scope of the present study. Nevertheless, we make a few preliminary remarks here, deferring a detailed investigation to future work. 
 
 On general grounds, we expect the rate for monopole-induced $(B+L)$-violation to be governed by an equation of the form
\begin{equation}
\label{eq:mon1}
\frac{ d n_{B+L}}{dt} = - \frac{\Gamma_\mathrm{M}}{V T^3} n_{B+L} + \cdots
\end{equation}
where $n_{B+L}$ is the $B+L$ density, $\Gamma_\mathrm{M}$ is the rate for monopole-catalyzed $(B+L)$-violation, and the $+\cdots$ indicate other ``collision terms" that we neglect for simplicity. Note that we have also neglected the impact of the expansion rate of the universe. Assuming an equilibrium monopole density as above, we then expect
\begin{equation}
\label{eq:mon2}
\Gamma_\mathrm{mon} = A_\mathrm{mon}(T) e^{-m_\text{M}(T)/T}\ \ \ ,
\end{equation}
where $A_\mathrm{mon}(T)$ depends on details of the $(B+L)$-violating amplitude(s). Since $B-L$ is conserved, Eqs.~(\ref{eq:mon1},\ref{eq:mon2}) can be interpreted as governing the total time-dependence of baryon number density, $n_B$. 

Now, let 
\begin{equation}
\label{eq:mon3}
\frac{n_B(\Delta t_\Sigma)}{n_B(0)}> e^{-X_\Sigma}
\end{equation}
denote the requirement on the baryon asymmetry at the completion of the $\Sigma$ phase that occurs a time $\Delta t_\Sigma$ after its onset needed to yield the present asymmetry. The value of $X_\Sigma$ will depend on the details of baryon number production during the first step from $O\to \Sigma$, including new CPV interactions that we have not addressed here. Integrating (\ref{eq:mon1}) over the time $\Delta t_\Sigma$ and requiring (\ref{eq:mon3}) will yield a BNPC \cite{Patel:2011th} of the form
\begin{equation}
\label{eq:mon4}
\frac{4\pi B_\text{M}}{g} \frac{\bar{x}(T)}{T} - 3 \ln\frac{\bar{x}(T)}{T} > -\ln X_\Sigma - \ln\left(\frac{\Delta t_\Sigma}{t_H}\right) + \ln Z_\text{M}\ \ \ ,
\end{equation}
where $t_H$ is the Hubble time and $Z_M$ depends on the prefactor $A_\mathrm{mon}(T)$. Note that (\ref{eq:mon4}) is similar in form to the BNPC for conventional electroweak baryogenesis for which $B+L$ violation is driven by sphaleron processes.  Consequently, to the extent that $Z_M$ is similar in magnitude to the corresponding quantity for sphaleron-driven washout, we expect a similar requirement on $\bar{x}(T)/T$ as one does for conventional electroweak baryogenesis [see Eq.~(\ref{eq:BNPC_SM}) below].

 As the foregoing arguments are only semi-quantitative, we defer a detailed analysis of the BNPC to a future study of $\Gamma_\mathrm{mon}$. Nonetheless, these considerations indicate that lowering the temperature of the $O\to\Sigma$ transition will lead to more effective baryon number preservation during the $\Sigma$ phase, whether monopole production occurs via Kibble mechanism of thermal pair production.

\subsection{Second Step: Klinkhammer-Manton Sphaleron} 
The $H$ phase within the bubbles of the second step of the EWPT is identical to the electroweak phase of the minimal Standard Model, wherein Klinkhammer-Manton sphalerons \cite{Klinkhamer:1984di} are known to exist.  Under the assumption that a sizable baryon asymmetry has survived during the first step, the sphalerons would again threaten to further reduce the asymmetry in the second step.  The analysis of the BNPC at this stage parallels that within the minimal Standard Model, giving the requirement\cite{Patel:2011th}
\begin{multline}\label{eq:BNPC_SM}
\frac{4\pi B}{g}\frac{\bar{v}(T_h)}{T_h}-6\ln\frac{\bar{v}(T_h)}{T_h}>\\
-\ln X_H-\ln\bigg(\frac{\Delta t_\text{EW}}{t_H}\bigg)+\ln\mathcal{Z}+\hbar\ln\kappa\,\,,
\end{multline}
where the $B$ is the integration over the sphaleron radial profile,  $\mathcal{Z}$ and $\kappa$ appear in the sphaleron rate prefactor,  $\Delta t_\text{EW}$ is the duration of the second step, and $X_H$ is defined analogously to Eq.~(\ref{eq:mon3}).  The additional factor of three in front of the $\ln \bar{v}(T)/T$ arises from zero mode fluctuations around the sphaleron. 
The precise value for $X_H$ depends not only on the baryon number production in the first step, but also on the extent of reduction due to the monopoles.  In spirit of the previous subsection, we defer a detailed analysis to a future study, and tentatively adopt the bound 
\begin{equation}
\frac{\bar{v}(T_h)}{T_h}\gtrsim 1\,
\end{equation}
usually quoted in the literature as a rule-of-thumb in our analysis of the electroweak phase transition.

\section{EW phase transition}
\label{sec:EWPT}

In this section, we turn our attention to the dynamics of the EWPT in the $Z_2$$\Sigma$SM. Before doing so, we comment on the issue of gauge invariance that has been at topic of recent interest in this context. One of the unresolved theoretical issues plaguing perturbative analyses of the electroweak phase transition is the problem of extracting gauge-invariant quantities relevant to baryogenesis.  The root of the problem lies in the lack of a gauge-invariant definition of the free energy that is compatible with perturbation theory.  Although the problem has been known since the early days of the development of thermal gauge theories \cite{Dolan:1974gu,Nielsen:1975fs}, it is in our view that the problem has not yet been solved satisfactorily.  The choice of gauge most commonly employed in the literature is the 't Hooft background $R_\xi$ Landau $\xi=0$ gauge.

In an earlier paper by us \cite{Patel:2011th}, the issue was tackled, and a theoretically sound method to carry out an analysis that is fully gauge-independent and compatible with perturbation theory was developed.  While the method applied to the SM as well as some features of the minimal supersymmetric SM gives the correct qualitative dependence on model parameters when compared against lattice results, maintaining gauge-independence at the perturbative level appears to come at the expense of numerical accuracy.

With this caveat in mind, the attitude we adopt here is the moral one: to compute physical observables in a way that suffer from no gauge ambiguities, and so we will follow the procedure detailed in our earlier paper with the awareness of the method giving correct parametric dependence but not numerical accuracy. We now proceed using the gauge-independent procedure as outlined in \cite{Patel:2011th} for finite temperature analysis of the free energy of the system.

The construction of the finite temperature one-loop effective potential follows the standard background-field method \cite{Dolan:1973qd}.  We write the resulting effective potential schematically as:
\begin{align}\label{eq:fullpotential}
\nonumber V(h,\sigma,T)&=V_\text{tree}(h,\sigma)\\
&\qquad+\hbar\big(V_1^{T=0}(h,\sigma)+V_1^{T\neq 0}(h,\sigma,T)\big)\,,
\end{align}
where $V_\text{tree}(h,\sigma)$ is the tree-level potential in (\ref{eq:treePotential}), and $V_1^{T=0}(h,\sigma)$ and $V_1^{T\neq 0}(h,\sigma,T)$ are the zero-temperature Coleman-Weinberg and temperature-dependent potential.  Explicit expressions in the 't Hooft $R_\xi$ background-field gauge is given in appendix \ref{ap:potential}. 

We follow the evolution of the minima of the effective potential as a function of temperature by inserting the critical values of the tree-level potential defined by (\ref{eq:minCond}) into the one-loop temperature-dependent potential (\ref{eq:fullpotential}), at which point the gauge dependence cancels.  As already discussed above, there are generally four such critical values to follow, corresponding to phases at $H$, $\Sigma$, $O$, and $T$ in Figure \ref{fig:vacuumSketch}.  We determine critical temperatures by requiring that the degeneracy condition
\begin{equation}
V(h_c^{(1)},\sigma_c^{(1)},T_c)=V(h_c^{(2)},\sigma_c^{(2)},T_c)
\end{equation}
be met.

\begin{figure}
\includegraphics[scale=1.0,width=7.2cm, trim=1cm 0 1cm 0]{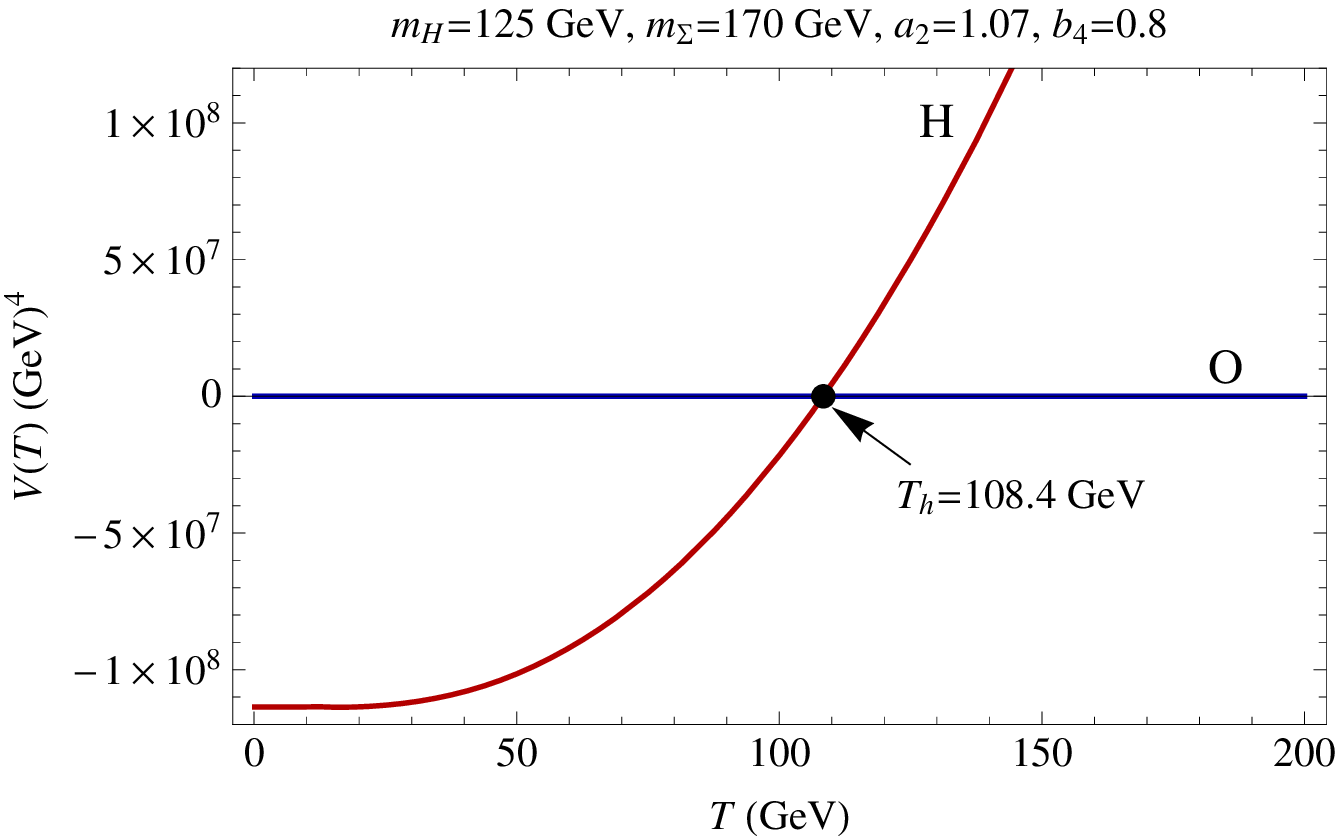}\\
\vspace{1cm}
\includegraphics[scale=1.0,width=7.2cm, trim=1cm 0 1cm 0]{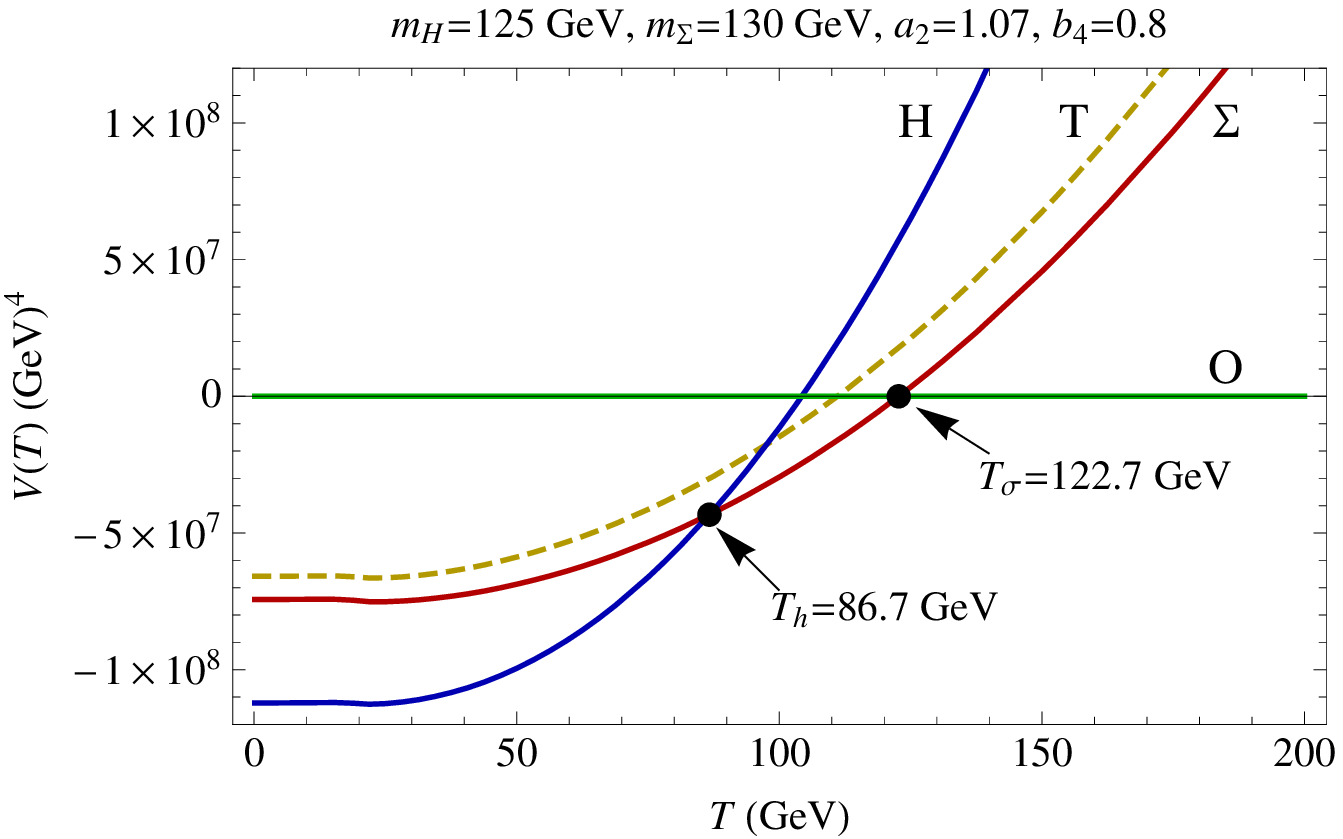}
\caption{Extremum points of the potential as a function of temperature for two choices of model parameters.  Curves are labeled according to phases defined in Fig. \ref{fig:vacuumSketch}.  Upper panel: the there is only one critical point, corresponding to a SM-like phase transition ($O\rightarrow H$); lower panel: the system exhibits two critical temperatures, favorable for baryogenesis, corresponding to transitions at critical temperatures $T_\sigma$: $O\rightarrow\Sigma$ and $T_h$: $\Sigma\rightarrow H$.}
\label{fig:freeEnergy}
\end{figure}

In Fig. \ref{fig:freeEnergy}, we provide two representative examples for the evolution of the free energy at its critical points as a function of temperature.  In Fig. \ref{fig:freeEnergy}a, we choose a large triplet scalar mass $m_\Sigma=170\text{ GeV}$, corresponding to a point inside the yellow region of Fig. \ref{fig:vacuumStab}a,
where the zero-temperature potential takes the form as in Fig. \ref{fig:potentials}a with no critical point along the $\sigma$ direction.  The evolution of the phase structure with temperature corresponds to a trajectory that follows the curve of lowest energy. Thus, at high-$T$ the universe evolves (from right to left) along the (blue) line labeled $O$, corresponding to the symmetric phase. For $T<T_h=108.4\text{ GeV}$, the (red) curve labeled $H$, giving the value of the potential at the electroweak minimum, has the lowest energy, so the trajectory then switches to the (red) $H$ curve.  A first order transition to the $H$ phase will then occur at $T$ just below $T_h$ if the nucleation probability is sufficiently large.  In the SM, this transition occurs via cross over rather than bubble nucleation.

On the other hand in Fig.~\ref{fig:potentials}b, for a lighter triplet mass $m_\Sigma=130\text{ GeV}$, corresponding to a point in the blue region of Fig.~\ref{fig:vacuumStab}a, the zero-temperature potential takes the form as in Fig.~\ref{fig:potentials}b, where a metastable minimum along the $\sigma$ direction exists.  Consequently, the system exhibits a richer two-step phase transition, favorable for baryogenesis, with critical temperatures at $T_\sigma=122.7\text{ GeV}$ and $T_\sigma=86.7\text{ GeV}$.
\begin{figure}
\includegraphics[scale=1.0,width=7cm]{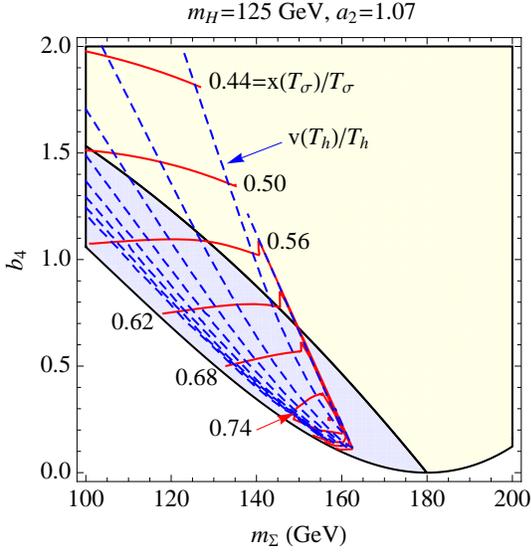}
\caption{Phase transition order parameters: Red contours indicate constant $\bar{x}(T_\sigma)/T_\sigma$ for first step.  For the second step, dashed blue contours correspond to $\bar{v}(T_h)/T_h$ with values \{1.2,\, 1.6,\, 2.0,\, 2.4,\, 2.8,\, 3.2,\, 3.6,\, 4.0\}, read right to left.  Outside the contoured region, the EWPT proceeds in a single SM-like step ($O\rightarrow H$) and is unfavorable for baryogenesis.}
\label{fig:paramScan1}
\end{figure}
The quantity relevant to the preservation of the baryon asymmetry at the time of the electroweak phase transition is the order parameter at the nucleation temperature: $\bar{x}(T^\text{nuc}_\sigma)/T^\text{nuc}_\sigma$ for the first step, and  $\bar{v}(T^\text{nuc}_h)/T^\text{nuc}_h$ for the second step, as defined above.  We use the more conservative order parameter, evaluated at the critical temperature, lying somewhat higher than the nucleation temperature.

In Fig. \ref{fig:paramScan1}, we display the results of a scan for fixed $a_2=1.07$ and $m_H=125\text{ GeV}$, which is superimposed over the tree-level vacuum stability graph of Fig. \ref{fig:vacuumStab}.  The contours are shown only where a two-step phase transition occurs.  The red contours correspond to lines of constant order parameter $\bar{x}(T_\sigma)/T_\sigma$, relevant for the first step, and the blue contours are lines of constant  $\bar{v}(T_h)/T_h$ for the second step. We observe that the strength of the transition in the first step, characterized by ${\bar x}(T_\sigma)/T_\sigma$, is relatively insensitive to $m_\Sigma$ for fixed $a_2$, but increases monotonically with decreasing $b_4$. This situation is analogous to the  pure SM case where baryon number preservation becomes more effective for smaller Higgs quartic self-coupling. Unlike the SM, however, the $\Sigma^0$ mass in the $H$ phase does not depend on its quartic self-coupling, so one may vary $b_4$ at will without encountering phenomenological constraints associated with $m_\Sigma$.

The value of $m_\Sigma$ does, however, affect the strength of the second step. Importantly, we find that wherever the two-step phase transition occurs in the parameter space, the second step of the phase transition is always strongly first order and comfortably satisfies the required BNPC in the Higgs direction.  As one moves towards lower values of $m_\Sigma$ holding $a_2$ and $b_4$ fixed (approaching the left edge of the blue region in Fig. \ref{fig:paramScan1}), the order parameter $\bar{v}(T_h)/T_h$ of the second step of the phase transition dramatically rises.  However, this raises the issue of whether substantial supercooling of the $\Sigma$-phase would occur leading to significant entropy injection, effectively diluting the baryon asymmetry through reheating; or if the second phase transition ever completes at all.

\begin{figure}
\includegraphics[width=7.0cm, trim=5.5cm 0 5.5cm 0]{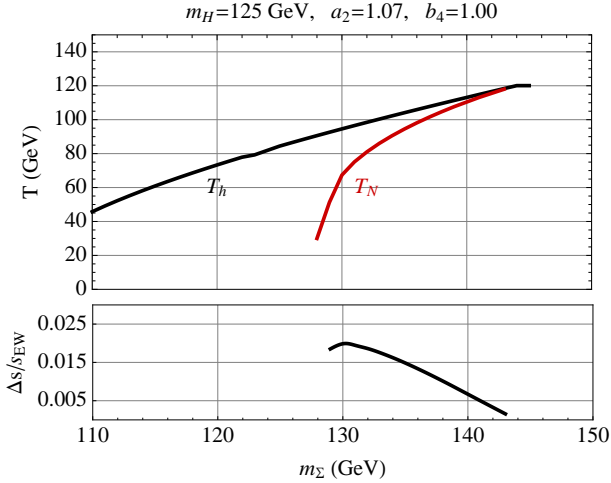}\\
\caption{Critical temperature (black), nucleation temperature (red) and entropy dilution factor $\Delta s/s_\mathrm{EW}$ as a function of triplet scalar mass $m_\Sigma$ for the second step of the phase transition ($\Sigma\rightarrow H$).  For light triplet masses, the phase transition never occurs (no nucleation temperature), as the universe is locked into the meta-stable $\Sigma$-phase at zero temperature. Hence, the curves for $T_N$ and entropy dilution terminate. }
\label{fig:entropy}
\end{figure}

To estimate the extent to which supercooling occurs, we carry out a numerical calculation of the nucleation rate in the semi-classical approximation.  We use a modified version of CosmoTransitions software \cite{Wainwright:2011kj} which rapidly solves for the `critical bubble' in theories with multiple field directions\footnote{A critical bubble is one that continues to expand after formation.}.  In spirit of maintaining gauge-independence, the code was modified to analyze just the $\mathcal{O}(T^2)$ part of the high-$T$ expansion of the thermal potential (\ref{eq:fullpotential}):
\begin{align}\label{eq:highT_V}
\nonumber V(h,s;\,T)&=D_h(T^2-T_{0h}^2)h^2+D_\sigma(T^2-T_{0\sigma}^2)\sigma^2\\
&\qquad+\frac{1}{4}(\lambda_h h^4+a_2 h^2\sigma^2 + b_4 \sigma^4).
\end{align}
Here the coefficients
\begin{align*}
D_h&=\frac{1}{32}(8\lambda+g'^2+3 g^2+4y_t^2+2a_2)\\
D_\sigma&=\frac{1}{24}(2a_2+5b_4+6g_2^2)\\
T_{0h}^2&=\frac{\mu^2}{2D_h}\\
T_{0\sigma}^2&=\frac{\mu_\Sigma^2}{2D_\sigma}\,.
\end{align*}
are all gauge-independent.  Once the nucleation temperature is found, we derived the entropy density injection from the free energy:
\begin{equation}
\Delta s=-\left(\frac{d V}{d T}\bigg|_\Sigma-\frac{d V}{d T}\bigg|_H\right)\,.
\end{equation}
normalized by the entropy density at the time of the electroweak phase transition
\begin{equation}
s_\text{EW}=\frac{2\pi^2}{45}g_{*s}T_\text{EW}^3\,,
\end{equation}
where we take $g_{*s}=100$.  Numerical results are shown in Fig. \ref{fig:entropy}, for two choices of $b_4$, and fixed $m_H$ and $a_2$.  The upper sections of each plot displays the critical temperature $T_h$ (black curve) and the nucleation temperature $T_N$ (red curve) as a function of the triplet mass $m_\Sigma$.  As the triplet mass $m_\Sigma$ is parametrically lowered from high values, the $\Sigma$ phase becomes more supercooled and the onset of the second phase transition is delayed.  Below a critical value for $m_\Sigma$, the second step of the phase transition never occurs, marked by the end of the red curve; the system remains in the metastable $\Sigma$ phase at zero temperature\footnote{ Note that a critical temperature $T_h$ may still exist, indicating degeneracy of the two minima, even if a transition from one to the other does not actually occur.}.

In the lower section of each plot, the entropy dilution factor $\Delta s/s_\text{EW}$ is shown.  Entropy dilution typically does not exceed 3\%, which is lies within the typical limits of numerical uncertainty for baryogenesis computations.

\section{Connection to Collider Phenomenology}
\label{sec:pheno}
Because the dynamics of the two-step transition depend on the interaction between the isospin triplet and Higgs doublet, it is interesting to ask how the viability of this scenario is affected by the recent observation of a bosonic resonance near 125 GeV \cite{aad:2012gk, chat:2012gu} that may be identified by the SM Higgs.  Of particular relevance  are results for the $H\rightarrow\gamma\gamma$ decay channel, for which the ATLAS \cite{ATLAS:2013oma} collaboration reports a marginally significant excess over the expected Standard Model signal strength while the CMS \cite{CMS:ril} results are consistent with the Standard Model expectations.
As discussed in \cite{FileviezPerez:2008bj}, the coupling of the Higgs doublet to the real triplet through $a_2$ modifies the branching fraction to two photons since the charged components contribute to the rate at the one-loop level.  In this section we discuss the connection of the $a_2$ parameter to the electroweak phase transition.

The $H\rightarrow\gamma\gamma$ branching fraction in the Standard Model is dominated by contributions from the $W$ gauge boson and top quark loops\cite{Djouadi:2005gi}:
\begin{equation}\label{eq:SMdecay}
\Gamma_{H\rightarrow\gamma\gamma}^\text{SM}=\frac{1}{4\pi}\frac{1}{m_H}\left|g_W F_{1}\big(\textstyle\frac{m_H^2}{m_W^2}\big)+g_t F_{1/2}\big(\textstyle\frac{m_H^2}{m_t^2}\big)\right|^2
\end{equation}
The $W$ boson gives a positive real contribution, whereas the top quark gives negative real contribution.  Near the observed Higgs mass of $m_H\approx 125$ GeV, the two processes interfere destructively, with the $W$ boson contribution dominating that of the top quark.

\begin{figure}
\includegraphics[scale=1.0,width=7.0cm,trim=0.5cm 0 0.5cm 0]{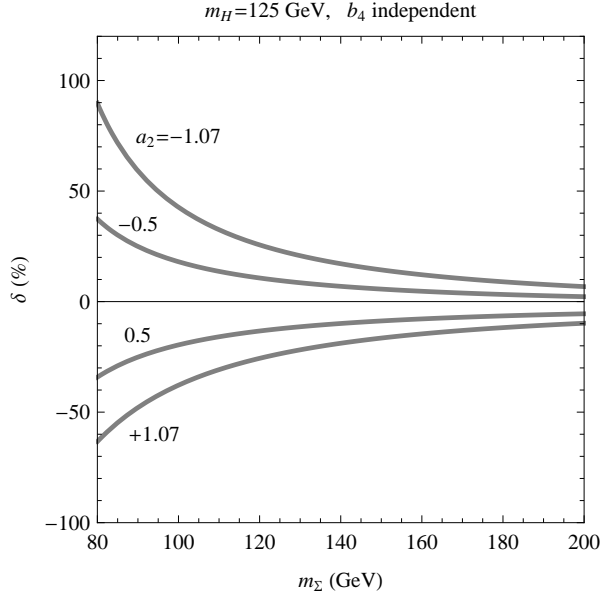}
\caption{Predictions for the $H\rightarrow\gamma\gamma$ branching fraction shift $\delta$ as defined in (\ref{eq:enhanceFactor}), as a function of triplet mass $m_\Sigma$ for select values of H-$\Sigma$ coupling $a_2$.  There is enhancement for negative $a_2$ and suppression for positive $a_2$.}
\label{fig:enhanceFactor}
\end{figure}

The presence of the charged triplet in this model modifies\cite{Gunion:1989we} the decay formula by the addition of the scalar contribution to the amplitude in (\ref{eq:SMdecay}):
\begin{align}
\nonumber \Gamma_{H\rightarrow\gamma\gamma}^\text{$\Sigma$ SM}&=\frac{1}{4\pi}\frac{1}{m_H}\bigg|g_W F_{1}\big(\textstyle\frac{m_H^2}{m_W^2}\big)\\
&\qquad+g_t F_{1/2}\big(\textstyle\frac{m_H^2}{m_t^2}\big)+g_\Sigma F_0(\textstyle\frac{m_H^2}{m_\Sigma^2})\bigg|^2\,,
\end{align}
where $g_\Sigma=e^2 v_0 a_2/(4\pi^2)$ is the effective coupling, independent of $b_4$.  The loop function $F_0(x)$ is real and negative for $m_\Sigma$ below threshold $x<4$, leading to an additional destructive interference against the $W$ contribution for positive $a_2$, and constructive interference for negative $a_2$.  We define the $H\rightarrow\gamma\gamma$ branching fraction shift relative to the Standard Model prediction
\begin{equation}\label{eq:enhanceFactor}
\delta=\frac{\Gamma_{H\rightarrow\gamma\gamma}^\text{$\Sigma$ SM}-\Gamma_{H\rightarrow\gamma\gamma}^\text{SM}}{\Gamma_{H\rightarrow\gamma\gamma}^\text{SM}}\,,
\end{equation}
and illustrate the effect of the presence of the charged scalar for specific values of $a_2$ in Fig. \ref{fig:enhanceFactor}.
\begin{figure}
\includegraphics[scale=1.0,width=7.0cm, trim=1.0cm 0 1.0cm 0]{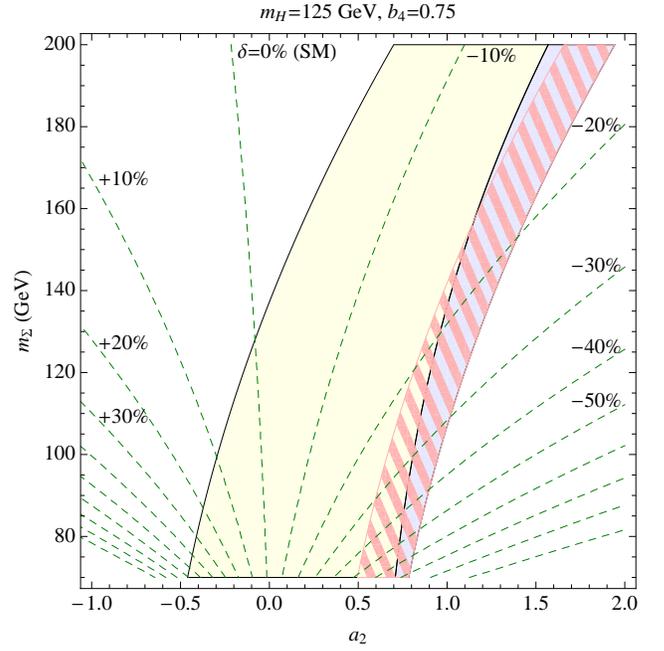}\\
\caption{Contours of constant $H\rightarrow\gamma\gamma$ branching fraction shift $\delta$  defined in (\ref{eq:enhanceFactor}), superimposed on tree-level vacuum stability plot.  The yellow and blue regions are color coded in the same way as in Fig. \ref{fig:vacuumStab}, and the hashed red area is the region where a two-step phase transition favorable for baryogenesis is expected.  Note the strong correlation between the sign of $a_2$, $\delta$.}
\label{fig:twoPhoton}
\end{figure}
Reports from ATLAS and CMS both mildly favor $\delta > 0$. 

In Fig. \ref{fig:twoPhoton}, for fixed $b_4$ and $m_H$, we show the regions in parameter space satisfying tree-level vacuum stability with the same blue and yellow color-coding used in Fig. \ref{fig:vacuumStab}.  The red hashed area corresponds to the region in parameter space where the desired two-step phase transition is expected to occur.  We then superimpose contours of constant $\delta$ of (\ref{eq:enhanceFactor}).  Two key points emerge: (1) regions where the two-photon branching fraction is enhanced occurs for negative $a_2$, where the EW vacuum is unstable already at tree-level.  (2) Regions where a two-step phase transition is likely to occur coincides with a reduction of the two-photon branching fraction relative to the Standard Model.  However, the size of this reduction can be lessened for larger triplet scalar mass $m_\Sigma$ within the region where a two-step phase transition occurs.

 The CMS results \cite{CMS:ril} would be consistent with a reduction in $\Gamma(H\to\gamma\gamma)$ as implied by the baryogenesis-favorable two step transition parameter space. On the other hand, it is interesting to ask how one might alleviate the tension with the ATLAS reported excess \cite{ATLAS:2013oma}, should the latter persist with the advent of more data and updated analyses.
Doing so could be achieved in a number of ways. The presence of additional charged degrees of freedom, such as new fermions, could compensate for the $\Sigma^\pm$ loop contributions. Alternately, one might imagine a two-step transition arising for $a_2<0$. In the present minimal extension, doing so is not achievable since it would imply a tachyonic mass for $\Sigma^0$ and $\Sigma^\pm$ [see (\ref{eq:msigma})]. But by suitably coupling the ${\vec\Sigma}$ to a new scalar singlet $S$ through the operator $\frac{1}{2}c_2 S^2 \vec\Sigma\cdot\vec\Sigma$, while allowing $S$ to obtain a zero-temperature vev $x_S$, the new mass relation for $\Sigma^0$
\be
m_\Sigma^2=-\mu_\Sigma^2+\frac{1}{2}a_2 v_0^2+c_1x_S^2
\ee
could yield $m_\Sigma^2>0$ for $a_2<0$ and $c_2>0$. The viability of the two-step transition in this case will be the subject of forthcoming work.

\section{Discussion and Conclusions}

The observation of a potentially fundamental scalar at the LHC makes the paradigm of scalar field-driven symmetry breaking in the early universe more realistic than ever. It is then interesting to explore possible patterns of symmetry-breaking that could arise in the presence of additional scalar fields, particularly if these fields and the associated phase transition dynamics can help account for the origin of the visible and dark matter of the universe. In this study, we have analyzed a simple SM scalar sector extension that gives a prototype for a multi-step EWPT wherein the baryon asymmetry may be generated prior to the final transition to the SM EW vacuum while yielding a contribution to the dark matter relic density.  This $Z_2\Sigma$SM does not provide a complete solution to the baryogenesis and dark matter problems, as it lacks additional sources of CP-violation required for baryogenesis and since for a sub-TeV mass for the new scalar, only a fraction of the dark matter relic density can be achieved. Nonetheless, it illustrates some of the generic features that may be present in more complete scenarios of this type: 
\begin{itemize}
\item A strong first EWPT during an initial step in which bubble nucleation can occur 
\item Suppression of monopole-catalyzed $B+L$ violation during this step that results from the coupling of the new scalars to the gauge sector of the SM and that is relatively insensitive to the mass of the new scalar
\item Subsequent transition to the EW vacuum that does not re-activate the SM sphalerons or lead to dangerous entropy injection and whose character is governed by the \lq\lq Higgs portal" scalar operator
\item The presence of a discrete symmetry that ensures stability of the neutral component of the new scalar multiplet in the EW phase, making it a contributor to the dark matter relic density.
\end{itemize}
Many, if not all, of these features can be realized in SM extensions with higher-dimensional scalar representations, some of which may contain new CP-violating phases in the scalar potential (or through couplings to fermions) as needed for baryogenesis and/or additional dark matter candidates.  

From a phenomenological perspective, it is also an interesting time to explore the possibilities for realizing this scenario. As we found in this work,  measurements of the $H\to\gamma\gamma$ signal strength provide an important test of our prototype scenario. As of this writing the ATLAS and CMS results are inconclusive, though the situation may change with the collection and analysis of additional data.  
More generally, the  presence of additional scalar states would lead to novel collider signatures that may be explored with increased luminosity and higher energy at the LHC.

\begin{acknowledgements}
The authors thank Max Wainwright for helpful correspondence regarding the CosmoTransitions software used for part of this work.  HP also thanks Andrew Long for interesting discussions.  This work was supported in part by the U.S. Department of Energy contract  DE-FG02-08ER41531 and by the Wisconsin Alumni Research Foundation.
\end{acknowledgements}

\begin{widetext}
\appendix
\section{Explicit form of the effective potential}\label{ap:potential}
The zero temperature (Coleman-Weinberg) and finite temperature parts of the 1-loop effective potentials used in the phase transition analysis is given here.  The dependence on the gauge parameter $\xi$ is explicitly shown.
\begin{multline}\label{eq:TindepVeff}
V_1^{T=0}(h)=\frac{1}{4(4\pi)^2}(m_H^2)^2\big[\ln(\textstyle\frac{m_H^2}{\mu^2})-\frac{3}{2}\displaystyle\big]+\frac{1}{4(4\pi)^2}(m_\Sigma^2)^2\big[\ln(\textstyle\frac{m_\Sigma^2}{\mu^2})-\frac{3}{2}\displaystyle\big]
+\frac{2\times1}{4(4\pi)^2}(m_\pm^2+\xi m_W^2)^2\big[\ln(\textstyle\frac{m_\pm^2+\xi m_W^2}{\mu^2})-\frac{3}{2}\displaystyle\big]\\
+\frac{1}{4(4\pi)^2}(m_G^2+\xi m_Z^2)^2\big[\ln(\textstyle\frac{m_G^2+\xi m_Z^2}{\mu^2})-\frac{3}{2}\displaystyle\big]
+\frac{2\times3}{4(4\pi)^2}(m_W^2)^2\big[\ln(\textstyle\frac{m_W^2}{\mu^2})-\frac{5}{6}\displaystyle\big]
+\frac{3}{4(4\pi)^2}(m_Z^2)^2\big[\ln(\textstyle\frac{m_Z^2}{\mu^2})-\frac{5}{6}\displaystyle\big]
\\
-\frac{2\times1}{4(4\pi)^2}(\xi m_W^2)^2\big[\ln(\textstyle\frac{\xi m_W^2}{\mu^2})-\frac{3}{2}\displaystyle\big]
-\frac{1}{4(4\pi)^2}(\xi m_Z^2)^2\big[\ln(\textstyle\frac{\xi m_Z^2}{\mu^2})-\frac{3}{2}\displaystyle\big]-\frac{4\times3}{4(4\pi)^2}(m_t^2)^2\big[\textstyle\ln(\frac{m_t^2}{\mu^2})-\frac{3}{2}\big]-\text{``free''}\,,
\end{multline}
and
\begin{multline}\label{eq:TdepVeff}
V^{T\neq0}_1(h,T)=\frac{T^4}{2\pi^2}\bigg[J_B\Big(\frac{m_H^2}{T^2}\Big)+J_B\Big(\frac{m_\Sigma^2}{T^2}\Big)+2\!\times\!J_B\Big(\frac{m_\pm^2+\xi m_W^2}{T^2}\Big)+J_B\Big(\frac{m_G^2+\xi m_Z^2}{T^2}\Big)\bigg]\\
+\frac{3T^4}{2\pi^2}\bigg[2\!\times\!J_B\Big(\frac{m_W^2}{T^2}\Big)+J_B\Big(\frac{m_Z^2}{T^2}\Big)+J_B\Big(\frac{m_\gamma^2}{T^2}\Big)\bigg]
-\frac{T^4}{2\pi^2}\bigg[2\!\times\!J_B\Big(\frac{\xi m_W^2}{T^2}\Big)+J_B\Big(\frac{\xi m_Z^2}{T^2}\Big)+J_B\Big(\frac{\xi m_\gamma^2}{T^2}\Big)\bigg]\\
-\frac{4T^4}{2\pi^2}\bigg[3\times J_F\Big(\frac{m_t^2}{T^2}\Big)\bigg]-\text{``free''}\,,
\end{multline}
where ``free'' represents a free-field subtraction.

Here, the field dependent scalar masses are given by:
\begin{align*}
m_H^2(h,\sigma)&=-\mu^2+3\lambda h^2+\frac{1}{2}a_2\sigma^2\\
m_\Sigma^2(h,\sigma)&=-\mu_\Sigma^2+\frac{1}{2}a_2 h^2+3 b_4 \sigma^2\\
m_G^2(h,\sigma)+\xi m_Z^2(h,\sigma)&=-\mu+\lambda h^2+\frac{1}{2}a_2\sigma^2+\frac{\xi}{4}(g^2+g'^2)h^2
\end{align*}
and the eigenvalues of
\begin{align*}
m_\pm^2(h,\sigma)+\xi m_W^2(h,\sigma)&=
\begin{pmatrix}
-\mu^2+\lambda h^2+\frac{a_2}{2}\sigma^2&0\\
0&-\mu_\Sigma^2+\frac{a_2}{2}h^2+b_4\sigma^2
\end{pmatrix}+\xi
\begin{pmatrix}
\frac{1}{4}g^2h^2&-\frac{1}{2}g^2h\sigma\\
-\frac{1}{2}g^2 h\sigma&g^2\sigma^2
\end{pmatrix}\,.
\end{align*}
The field dependent gauge boson masses are given by
\begin{align*}
m_W^2(h,\sigma)&=\frac{1}{4}g^2h^2+g^2\sigma^2\\
m_Z^2(h,\sigma)&=\frac{1}{4}(g^2+g'^2)h^2\\
m_\gamma^2(h,\sigma)&=0
\end{align*}
and among the SM fermions, we include only the top-quark, whose field-dependent mass given by
\begin{equation}
m_t^2(h,\sigma)=\frac{1}{2}y_t^2 h^2.
\end{equation}
\end{widetext}

\bibliography{SigmaSM.bib}

\end{document}